\documentstyle[12pt]{article}
\input epsf.sty
\renewcommand{\baselinestretch}{1.0}

 1

\catcode`\@=11 
\makeatletter
\def\@seccntformat#1{\csname the#1\endcsname.\hskip 1em}

\makeatother
\begin{document}

\newcommand{\as}{\alpha_s}
\newcommand{\A}{{\cal A}}
\newcommand{\B}{{\cal B}}
\newcommand{\Oa}{{\cal O}(\alpha_s)}
\newcommand{\Oaa}{{\cal O}(\alpha_s^2)}
\newcommand{\Oaaa}{{\cal O}(\alpha_s^3)}
\newcommand{\lam}{\Lambda_{\overline{MS}}}
\newcommand{\aspi}{\tilde{\alpha}_s}
\newcommand{\z}{$Z^0$}
\newcommand{\zcc}{$Z^0 \rightarrow c\bar{c}$}
\newcommand{\zbb}{$Z^0 \rightarrow b\bar{b}$}
\newcommand{\zuds}{$Z^0 \rightarrow u\bar{u},d\bar{d},s\bar{s}$}
\newcommand{\asz}{\alpha_s(M_Z^2)}
\newcommand{\llb}{$\Lambda^0/\bar{\Lambda}^0$}
\newcommand{\pnb}{differential cross section}
\newcommand{\pnc}{production rate}
\newcommand{\pnd}{spectrum}

\thispagestyle{empty}
\begin{flushright}

{\footnotesize\renewcommand{\baselinestretch}{.75}
              SLAC-PUB-8159\\
                  June 1999\\
}
\end{flushright}

\begin{center}
{\large \bf  Production of Charged ${\bf \pi^{\pm}}$, ${\bf K^{\pm}}$
and p/$\bar{\rm {\bf p}}$ in Hadronic ${\bf Z^0}$ Decays}

\vspace {1.0cm}
\baselineskip=18pt 

 {\bf The SLD Collaboration$^{**}$}\\
Stanford Linear Accelerator Center \\
Stanford University, Stanford, CA~94309

\vspace{2.5cm}

\end{center}

\normalsize


\begin{center}
{\bf ABSTRACT }
\end{center}

We have updated our results on identified charged hadron production using the
full SLD data sample of 550,000 hadronic $Z^0$ decays taken between 1993 and
1998.
The SLD Cherenkov Ring Imaging Detector allows the identification of clean
samples of charged pions, kaons and protons over a wide momentum range,
providing precise tests of perturbative QCD calculations and of fragmentation
models.
We have studied flavor-inclusive $Z^0$ decays, as well as decays into light,
$c$ and $b$ flavors, selected using the SLD vertex detector.
In addition we have updated our comparison of hadron and antihadron production
in light quark (rather than antiquark) jets, selected using the high SLC
electron beam polarization.
Differences between hadron and antihadron production at high momentum fraction
provide precise measurements of leading particle production and new, stringent
tests of fragmentation models.

\vfil

\noindent
Contributed to:  the
International Europhysics Conference on High Energy Physics,
15--21 July 1999, Tampere, Finland; Ref: 1\_186, \\
and to the XIX$^{th}$ International Symposium on Lepton Photon Interactions,
August 9--14, 1999, Stanford, USA.

\vfil
 
\noindent
$^*$This work was supported by Department of Energy
  contract DE-AC03-76SF00515.

\eject

\renewcommand{\baselinestretch}{1.5}
\section{Introduction}

The production of final state hadrons from primary hard partons,
e.g. the quark and antiquark in $e^+e^- \rightarrow Z^0 \rightarrow q\bar{q}$,
is currently believed to proceed in three stages.
The first stage involves the radiation of gluons from the primary quark and
antiquark,
which in turn radiate gluons or split into $q\bar{q}$ pairs until their
virtuality approaches the hadron mass scale.
Such a ``parton shower" is
calculable in perturbative QCD, for example in the
Modified Leading Logarithm Approximation (MLLA)~\cite{mlla}.

The second stage, in which these partons turn into ``primary" hadrons, is not
understood quantitatively, although several hadronization models exist.
A simple model is the ansatz of Local Parton-Hadron Duality (LPHD)~\cite{mlla},
which hypothesizes that distributions of kinematic quantities for a given
hadron species are directly proportional to the parton distributions at some
appropriate parton virtuality.
This allows the prediction via MLLA QCD
of the shapes of \pnb s for primary hadrons,
and of, for example, the energy- and mass-dependences of the peak of the
distribution of $\xi=-\ln(x_p)$, where $x_p=2p/E_{cm}$,
$p$ is the hadron momentum
and $E_{cm}$ is the $e^+e^-$ center-of-mass energy.

The third stage, in which unstable primary hadrons decay into
final state hadrons, complicates the interpretation of inclusive
measurements.  It is desirable to remove the effects of these
decays when comparing with the predictions of QCD$+$LPHD.
Additional complications arise in jets initiated by heavy ($c$ or $b$) quarks
in which the leading heavy hadrons carry a large fraction of the beam energy,
restricting that available to other primary particles, and
then decay into a number of secondary particles.
It is thus also desirable to restrict measurements to events with light primary
flavors.
    

A particularly interesting aspect of jet fragmentation is the question of what
happens to the primary quark or antiquark that initiated the jet.
Many fragmentation models assume that the initial quark is ``contained" as a
valence constituent of a particular hadron, and that this ``leading"
hadron has on average a higher momentum than the other particles in the jet.
This phenomenon has not been studied precisely for high-energy light-flavor
jets, since it is difficult to
identify the sign and flavor of the initial $q$/$\bar{q}$ on a jet-by-jet basis.
The quantification of leading particle effects could lead to
ways to identify the primary flavor of arbitrary samples of jets,
enabling a number of new measurements in $e^+e^-$,
as well as in $e$p and p$\bar{\rm p}$, collisions.

In this paper we present an update of our analysis of $\pi^{\pm}$, $K^{\pm}$, 
and p/$\bar{\rm p}$ production in hadronic $Z^0$ decays collected by the
SLC Large Detector (SLD).
The analysis is based upon the full sample of 550,000 hadronic events obtained 
in runs of the SLAC Linear Collider (SLC) between 1993 and 1998.
We measure \pnb s
in an inclusive sample of hadronic events of all flavors,
and also in high-purity samples of light- (\zuds) and $b$-flavor (\zbb) events.
From these three samples we extract corrected \pnb s in light- and $b$-, as well
as $c$-flavor (\zcc) events.
The unfolded \pnb s for the light-flavor events are free from effects of heavy
quark production and decay, and as such provide a more appropriate sample for
comparison with QCD predictions, which generally assume massless quarks, 
although the influence of decay products of other unstable primary hadrons
remains.
We use these measurements to test the predictions of
various fragmentation models.

We also select samples of quark and antiquark jets from our light-flavor
event sample, using the large forward-backward production asymmetry in polar
angle inherent in collisions of highly polarized electrons with positrons.
The \pnb s
are measured separately for hadrons and antihadrons in light-quark jets, and the
observed differences are interpreted in terms of leading particle effects.
These measurements provide precise, unique tests of fragmentation models.

\section{The SLD and Hadronic Event Selection} 

A general description of the SLD can be found elsewhere~\cite{sld}.
The trigger and initial selection criteria for hadronic $Z^0$ decays are 
described in Ref.~\cite{sldalphas}.
This analysis used charged tracks measured in the Central Drift
Chamber (CDC)~\cite{cdc} and Vertex Detector (VXD)~\cite{vxd}, and identified
using the Cherenkov Ring Imaging Detector (CRID) \cite{crid}.
Momentum measurement is provided by a uniform axial magnetic field of 0.6T.
The CDC and VXD give a momentum resolution of
$\sigma_{p_{\perp}}/p_{\perp}$ = $0.01 \oplus 0.0026p_{\perp}$,
where $p_{\perp}$ is the track momentum transverse to the beam axis in
GeV/$c$.
One quarter of the data were taken with the original vertex detector (VXD2), and
the rest with the upgraded detector (VXD3).
In the plane normal to the beamline 
the centroid of the micron-sized SLC IP was reconstructed from tracks
in sets of approximately thirty sequential hadronic $Z^0$ decays to a precision 
of $\sigma_{IP}\simeq7$ $\mu$m for the VXD2 data and $\simeq$3 $\mu$m for the
VXD3 data.
Including the uncertainty on the IP position, the resolution on the 
charged track impact parameter ($\delta$) projected in the plane perpendicular
to the beamline is 
$\sigma_{\delta} =$11$\oplus$70/$(p \sin^{3/2}\theta)$ $\mu$m for VXD2 and
$\sigma_{\delta} =$8$\oplus$29/$(p \sin^{3/2}\theta)$ $\mu$m for VXD3, where
$\theta$ is the track polar angle with respect to the beamline. 
The CRID comprises two radiator systems that allow the identification of
charged pions
with high efficiency and purity in the momentum range 0.3--35 GeV/c, charged
kaons in the ranges 0.75--6 GeV/c and 9--35 GeV/c, and protons in the ranges
0.75--6 GeV/c and 10--46 GeV/c \cite{bfp}.
The event thrust axis~\cite{thrust} was calculated using energy clusters
measured in the Liquid Argon Calorimeter~\cite{lac}. 

A set of cuts was applied to the data to select well-measured tracks
and events well contained within the detector acceptance.
Charged tracks were required to have a distance of
closest approach transverse to the beam axis within 5 cm,
and within 10 cm along the axis from the measured IP,
as well as $|\cos \theta |< 0.80$, and $p_\perp > 0.15$ GeV/c.
Events were required to have a minimum of seven such tracks,
a thrust axis  polar angle w.r.t. the beamline, $\theta_T$,
within $|\cos\theta_T|<0.71$, and
a charged visible energy $E_{vis}$ of at least 20~GeV,
which was calculated from the selected tracks assigned the charged pion mass. 
The efficiency for selecting a well-contained $Z^0 \rightarrow q{\bar q}(g)$
event was estimated to be above 96\% independent of quark flavor.
The VXD, CDC and CRID were required to be operational, resulting in a 
selected sample of roughly 303,000 events, with an estimated
non-hadronic background contribution of $0.10 \pm 0.05\%$ dominated
by $Z^0 \rightarrow \tau^+\tau^-$ events.
 
Samples of events enriched in light and $b$ primary flavors were selected based
on charged track impact parameters $\delta$ with respect to the IP in the
plane transverse to the beam \cite{mikeh}.
For each event we define $n_{sig}$ as the number of tracks with
impact parameter greater than three times its estimated error,
$\delta > 3 \sigma_{\delta}$.
Events with $n_{sig}=0$
were assigned to the light flavor sample and those with $n_{sig} \geq 4$
were assigned to the $b$ sample; the remaining events were classified as a $c$
sample.
The light, $c$ and $b$ samples comprised 176,000, 88,000 and 38,000 events,
respectively;
selection efficiencies and sample purities were estimated from our
Monte Carlo simulation and are listed in table \ref{tlveff}.

\begin{table}
\begin{center}
 \begin{tabular}{|l||c|c|c||c|c|c|} \hline
       & \multicolumn{3}{c||}{ } & \multicolumn{3}{c|}{ } \\ [-.4cm]
       & \multicolumn{3}{c||}{Efficiency for $Z^0 \rightarrow$}
       & \multicolumn{3}{c|}{Purity of $Z^0 \rightarrow$} \\
       & $u\bar{u},d\bar{d},s\bar{s}$ & $c\bar{c}$ & $b\bar{b}$ 
       & $u\bar{u},d\bar{d},s\bar{s}$ & $c\bar{c}$ & $b\bar{b}$  \\ [.1cm] \hline 
 &&&&&&\\[-.4cm] 
light-tag  & 0.846 & 0.338 & 0.034 & 0.881 & 0.106 & 0.013 \\
$c$-tag    & 0.153 & 0.617 & 0.401 & 0.320 & 0.388 & 0.292 \\
$b$-tag    & 0.001 & 0.045 & 0.565 & 0.005 & 0.064 & 0.931 \\[.1cm] \hline 
 \end{tabular}
\caption{\baselineskip=12pt \label{tlveff}
Tagging efficiencies for simulated events in the three flavor categories to be
tagged as light, $c$ or $b$.  The
three rightmost columns indicate the composition of each simulated tagged sample
assuming SM relative flavor production.}
\end{center}
\end{table}

Separate samples of hemispheres enriched in light-quark and light-antiquark jets
were selected from the light-tagged event sample by exploiting the large
electroweak forward-backward production asymmetry wrt the beam direction.
The event thrust axis was used to approximate the initial $q\bar{q}$ axis and
was signed such that its $z$-component was positive, $\hat{t}_z>0$.
Events in the central region of the detector, where the production asymmetry is
small, were removed by the requirement $|\hat{t}_z|>0.2$, leaving
125,000 events.
The quark-tagged hemisphere in events with left-(right-)handed electron beam
was defined to comprise the set of tracks with positive (negative) momentum
projection along the signed thrust axis.
The remaining tracks in each event
were defined to be in the antiquark-tagged hemisphere.
The sign and magnitude of the electron beam polarization were measured for every
event.
For the selected event sample, the average magnitude of the 
polarization was 0.73.
Using this value and assuming Standard Model couplings at 
tree-level, the purity of the quark-tagged sample is 0.73.

For the purpose of estimating the efficiency and purity of the event
flavor tagging and the particle identification,
we made use of a detailed Monte Carlo (MC) simulation of the detector.
The JETSET 7.4~\cite{jetset} event generator was used, with parameter
values tuned to hadronic $e^+e^-$ annihilation data~\cite{tune},
combined with a simulation of $B$-hadron decays
tuned~\cite{sldsim} to $\Upsilon(4S)$ data and a simulation of the SLD
based on GEANT 3.21~\cite{geant}.
Inclusive distributions of single-particle and event-topology observables
in hadronic events were found to be well described by the
simulation~\cite{sldalphas}.


\section{Measurement of the Charged Hadron Fractions} 

Charged tracks were identified as pions, kaons or protons,
in the CRID using a likelihood technique \cite{davea}.
Information from the liquid (gas) radiator only was used for tracks with
$p<2.5$ ($p>7.5$) GeV/c; in the overlap region, $2.5<p<7.5$ GeV/c,
liquid and gas information was combined.
Additional track selection cuts were applied to remove tracks that 
scattered through large angles before exiting the CRID and 
to ensure that the CRID performance was well-modelled by the simulation.
Tracks were required to have at least 40 CDC hits, at least one of which
was in the outermost superlayer,
to extrapolate through an active region of the appropriate radiator(s),
and
to have at least 80 (100)\% of their expected liquid (gas) ring contained
within a sensitive region of the CRID TPCs.
The latter requirement included rejection of tracks with $p>2.5$ GeV/c for
which there was a saturated CRID hit (from passage of miminum-ionizing
particles) within a 5 cm radius (twice the maximum ring radius) of the expected
gas ring center.
Tracks with $p<7.5$ GeV/c were required to have
a saturated hit within 1 cm of the extrapolated track, and
tracks with $p>2.5$ GeV/c were required to have either such a saturated hit or
the presence of at least four hits consistent with a liquid ring.
These cuts accepted 47, 28 and 49\% of tracks within the barrel acceptance
in the momentum ranges $p<2.5$, $2.5<p<7.5$ and $p>7.5$ GeV/c, respectively.
For momenta below 2 GeV/c, only negatively charged tracks were used to reduce
the background from protons produced in interactions with the detector
material.
For momenta below 2.5 GeV/c, only the VXD2 data were used (due to time
constraints), and the results in this region are identical to our published
results \cite{bfp}.

\begin{figure}
 \hspace*{0.5cm}   
   \epsfxsize=4.5in
   \begin{center}\mbox{\epsffile{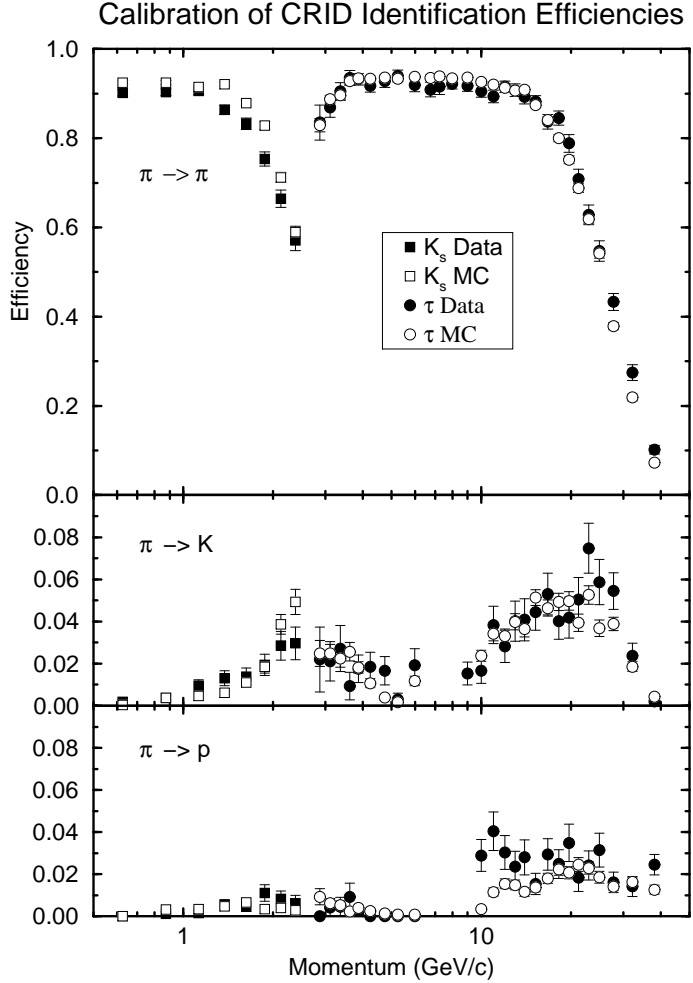}}\end{center}
  \caption{ 
 \label{ecal}
Calibration of the pion identification efficiencies using tracks from tagged
$K^0_s$ and $\tau^\pm$ decays.
    }
\end{figure}

\begin{figure}
 \hspace*{0.5cm}   
   \epsfxsize=5.9in
   \begin{center}\mbox{\epsffile{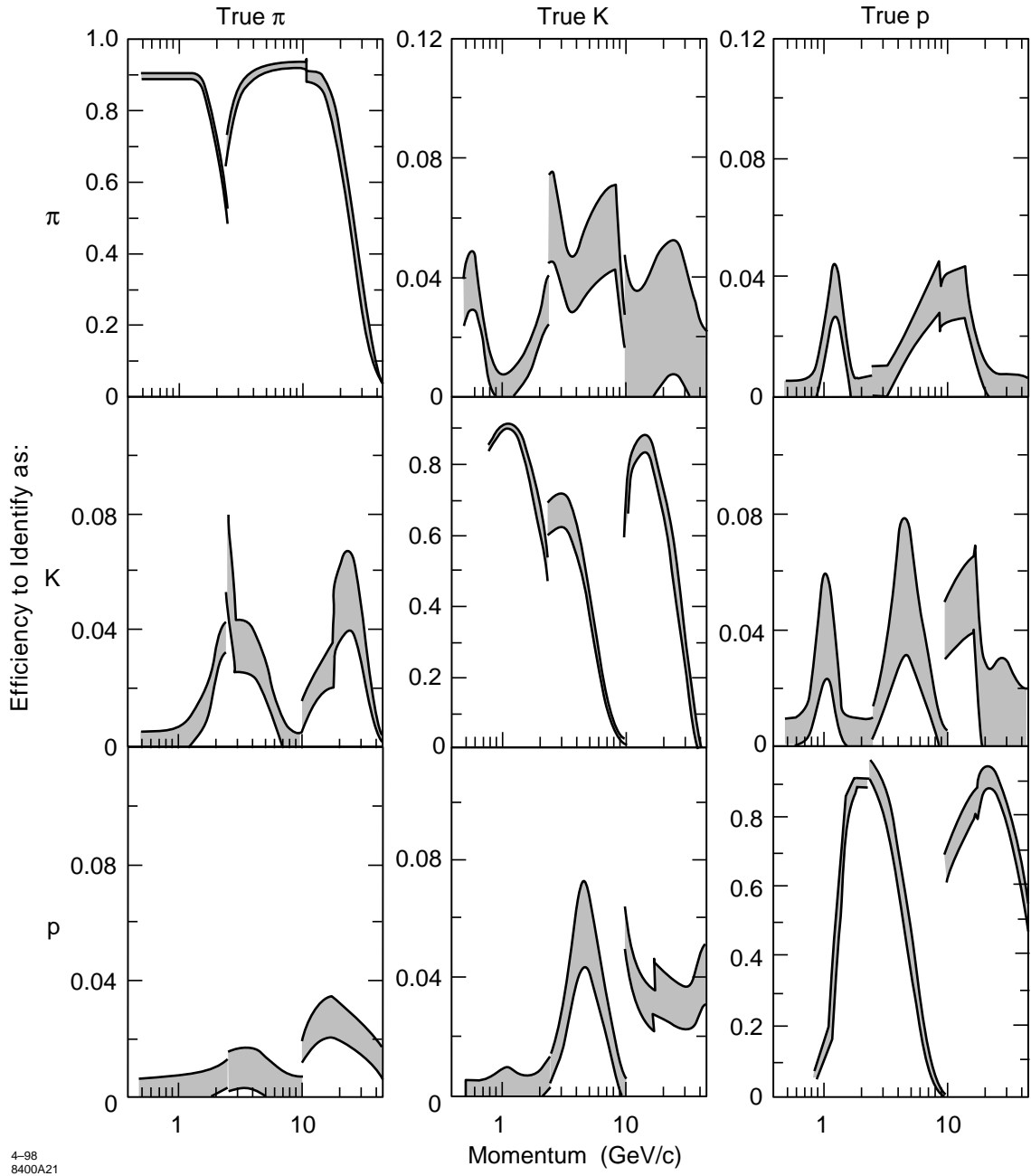}}\end{center}
  \caption{ 
 \label{effpar}
Calibrated identification efficiencies for tracks used in the charged hadron
fractions analysis.
The separations between the pairs of lines represent the systematic
uncertainties, which are strongly correlated between momenta.
 }
\end{figure}

For tracks with $p<2.5$ ($p>2.5$) GeV/c, we define a particle to be identified
as type $j$, where $j=\pi,K$,p, if ${\cal L}_j$
exceeds both of the other log-likelihoods by at least 5 (3) units.
Efficiencies for identifying selected particles of true type $i$
as type $j$ were determined where possible from the data, using tracks from
tagged $K^0_s$, $\tau^\pm$ and $\Lambda^0$ decays, as described in \cite{bfp}.
An example is shown in fig. \ref{ecal}.
A detailed Monte Carlo (MC) simulation of the detector was then used to make
small corrections to these measurements,
and to derive the remaining efficiencies from those measured.
These efficiencies are parametrized in terms of continuous functions in each of
the three momentum ranges, and are shown in
fig. \ref{effpar}, in which the pairs of lines represent our estimated
efficiencies plus and minus their systematic uncertainties.
For the diagonal entries, these uncertainties correspond to statistical errors
on the parameters fitted from the data, and are completely positively
correlated across each of the three momentum regions.
For the off-diagonal terms, representing misidentification rates, a more
conservative 25\% relative error was assigned at all points to account
for the limited experimental constraints on the momentum dependence.
These errors are also strongly positively correlated among momenta.
The diagonal elements peak near or above 0.9 and
the pion coverage is continuous from 0.5 GeV/c up to approximately 35 GeV/c.
There is a gap in the kaon-proton separation between 7 and 10 GeV/c due to
limited resolution of the liquid system and the fact that both particles are
below Cherenkov threshold in the gas system.
The proton coverage extends to the beam momentum.
Misidentification rates are typically less than 0.03, with peak values of
up to 0.07.

In each momentum bin we measured the fractions of the selected tracks
that were identified as $\pi$, $K$ and p.
The observed fractions were related to the true production fractions by an
efficiency matrix, composed of the values in fig.~\ref{effpar} for that bin.
This matrix was inverted and used to unfold our observed identified particle
rates.
This analysis procedure does not require that the sum 
of the charged particle fractions be unity; instead the sum was used as a 
consistency check and was found to be within statistical errors of unity for all
momenta. 
In some momentum regions we cannot distinguish two of the three species,
so the procedure was reduced to a 2$\times$2 matrix analysis and we present
only the fraction of the identified species, i.e. protons
above 35 GeV/c and pions between 6 and 9.5 GeV/c.

Electrons and muons were not distinguished from pions in this analysis;
this background was estimated from the simulation to be about
5\% in the inclusive flavor sample,
predominantly from $c$- and $b$-flavor events.
The flavor-inclusive fractions were corrected using the simulation for
the lepton backgrounds, as well as for the effects of beam-related
backgrounds, particles interacting in the detector material, and particles with
large flight distance, such that the conventional definition of a final-state
charged hadron is recovered,
namely charged pions, kaons or protons that are either
from the primary interaction or decay products of particles with lifetime less
than 3$\times10^{-10}$s.

\begin{figure}
 \hspace*{0.5cm}   
   \epsfxsize=6.2in
   \begin{center}\mbox{\epsffile{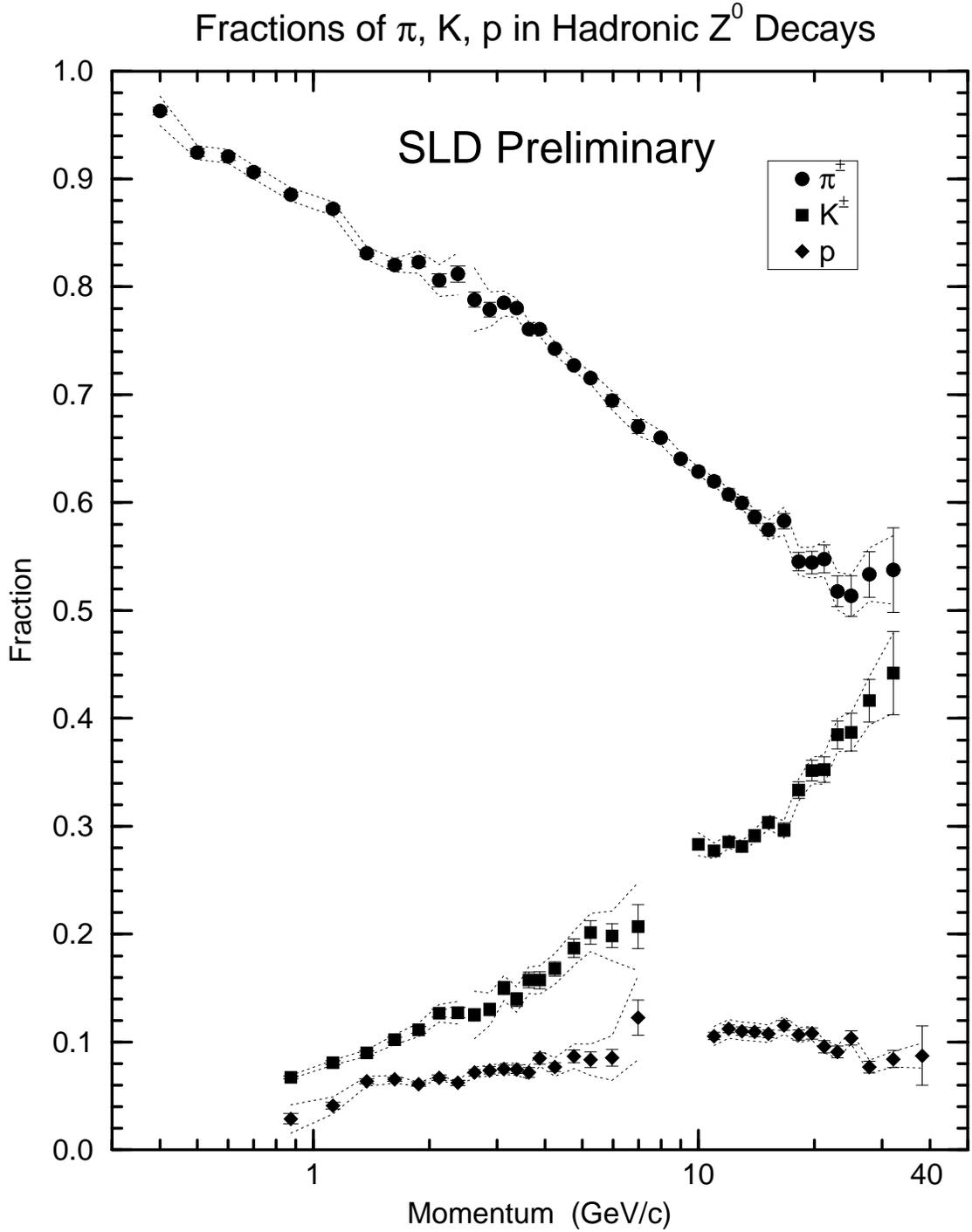}}\end{center}
  \caption{ 
 \label{fraxg}
Measured charged hadron production fractions in hadronic $Z^0$ decays. 
The circles represent the $\pi^\pm$ fraction, the squares the $K^\pm$ fraction,
the diamonds the p/$\bar{\rm p}$ fraction.
The error bars are statistical only.  The dotted lines indicate
the systematic errors, which are strongly correlated between
momenta (see text).
    }
\end{figure}

The measured charged particle fractions for inclusive hadronic $Z^0$ decays
are shown in fig.~\ref{fraxg}.
The errors on the points below 15 GeV/c are dominated by the systematic
uncertainties on the identification efficiencies and are
strongly positively correlated across the entire momentum range.
For $p>15$ GeV/c the errors have roughly equal statistical and systematic
contributions, and the systematic errors are positively correlated and
increase in magnitude with momentum.

Pions are seen to dominate the charged hadron production at low momentum,
and to decline steadily in fraction as momentum increases.
The kaon fraction rises steadily to about one-third at high momentum.
The proton fraction rises to a maximum of about one-tenth at about
10 GeV/c, then declines slowly.
At high $x_p$, the pion and kaon fractions appear to be converging.
This convergence could indicate reduced strangeness suppression at high
momentum, or that production is becoming dominated by leading particles,
such that kaons from $s\bar{s}$ events are as common as pions from $u\bar{u}$
and $d\bar{d}$ events.

Where the momentum coverage overlaps, these measured fractions were found to be 
in agreement with our previous results \cite{bfp} and with other measurements
at the $Z^0$~\cite{delphi,opal,aleph}.
Measurements based on ring imaging \cite{bfp,delphi} and those based on
ionization energy loss rates~\cite{opal,aleph} cover complementary
momentum ranges and can be combined to provide continuous
coverage over the range $0.2<p<35$ GeV/c.

\begin{figure}
\vspace{-1.cm}
 \hspace*{0.5cm}   
   \epsfxsize=6.5in
   \begin{center}\mbox{\epsffile{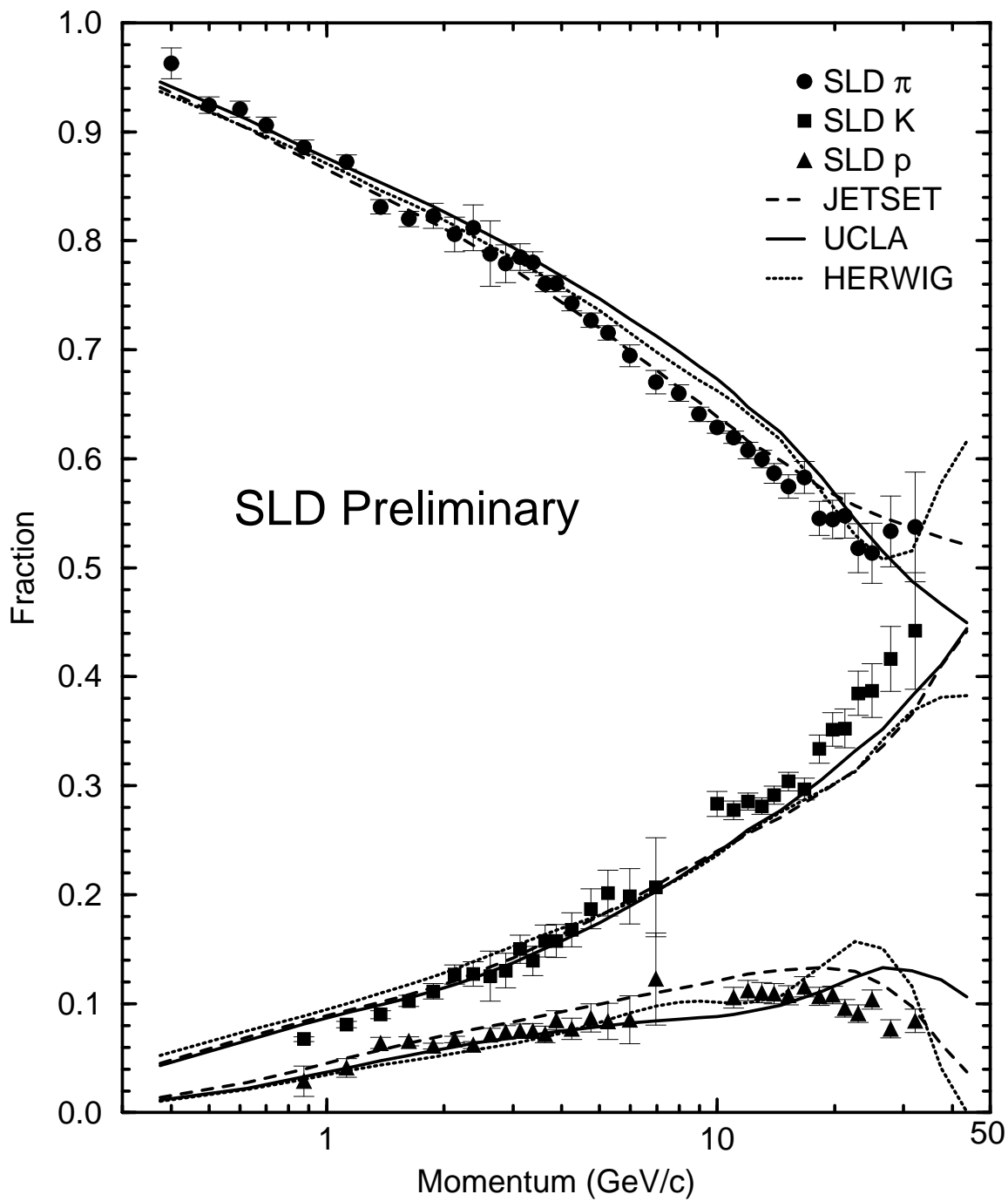}}\end{center}
  \caption{ 
 \label{fallmc}
Comparison of the charged hadron fractions in flavor-inclusive events with the
predictions of three fragmentation models.
    }
\end{figure}

In fig. \ref{fallmc} we compare our measured charged hadron fractions with the
predictions of the JETSET 7.4 \cite{jetset}, UCLA \cite{ucla} and
HERWIG 5.8 \cite{herwig} fragmentation models, using default parameters.
The momentum dependence of each fraction
is reproduced qualitatively by all three models.
The HERWIG and UCLA predictions for the pion fraction are high at intermediate
$x_p$;  the three model predictions differ widely at very high $x_p$, but the
statistics of the data are not sufficient to distinguish between them.
All three predictions for the kaon fraction are too low (high) at small
(large) $x_p$.
The JETSET prediction for the proton fraction is too high at all $x_p$;
those of HERWIG and UCLA show structure in the proton fraction at large $x_p$
that is inconsistent with the data.

\section{Flavor-Dependent Analysis}

The analysis was repeated separately on the high-purity light- and $b$-tagged
event samples described in section 2,
and on the remaining sample of events satisfying
neither tag requirement, which we denote $c$-tagged.
In each momentum bin the measured \pnb s $r^{meas}_j$ of each hadron species
for these three samples, $j=$light-tag, $c$-tag, $b$-tag,
were unfolded by inverting the relations:
\begin{equation}
r^{meas}_j =
\frac{\Sigma_i b_{ij} \epsilon_{ij} R_i r^{true}_i}{\Sigma_i \epsilon_{ij} R_i}
\end{equation}
to yield true \pnb s $r^{true}_i$ in events of the three flavor types,
$i=$1, 2, 3, corresponding to \zuds, \zcc\ and \zbb.
Here, $R_i$ is the fraction of hadronic $Z^0$ decays of flavor type $i$, taken
from \cite{pdg}, $\epsilon_{ij}$ is the event tagging efficiency matrix,
estimated from the simulation and listed in table \ref{tlveff},
and $b_{ij}$ represents the momentum-dependent bias of tag $j$ toward
selecting events of flavor $i$ that contain hadrons of the type in question.
The diagonal bias values \cite{tomp} are within a few percent of unity,
reflecting a small multiplicity dependence of the flavor tags.
The off-diagonal bias values are larger, but these have
little effect on the unfolded results.

\begin{figure}
\vspace{-1.cm}
 \hspace*{0.5cm}   
   \epsfxsize=6.4in
   \begin{center}\mbox{\epsffile{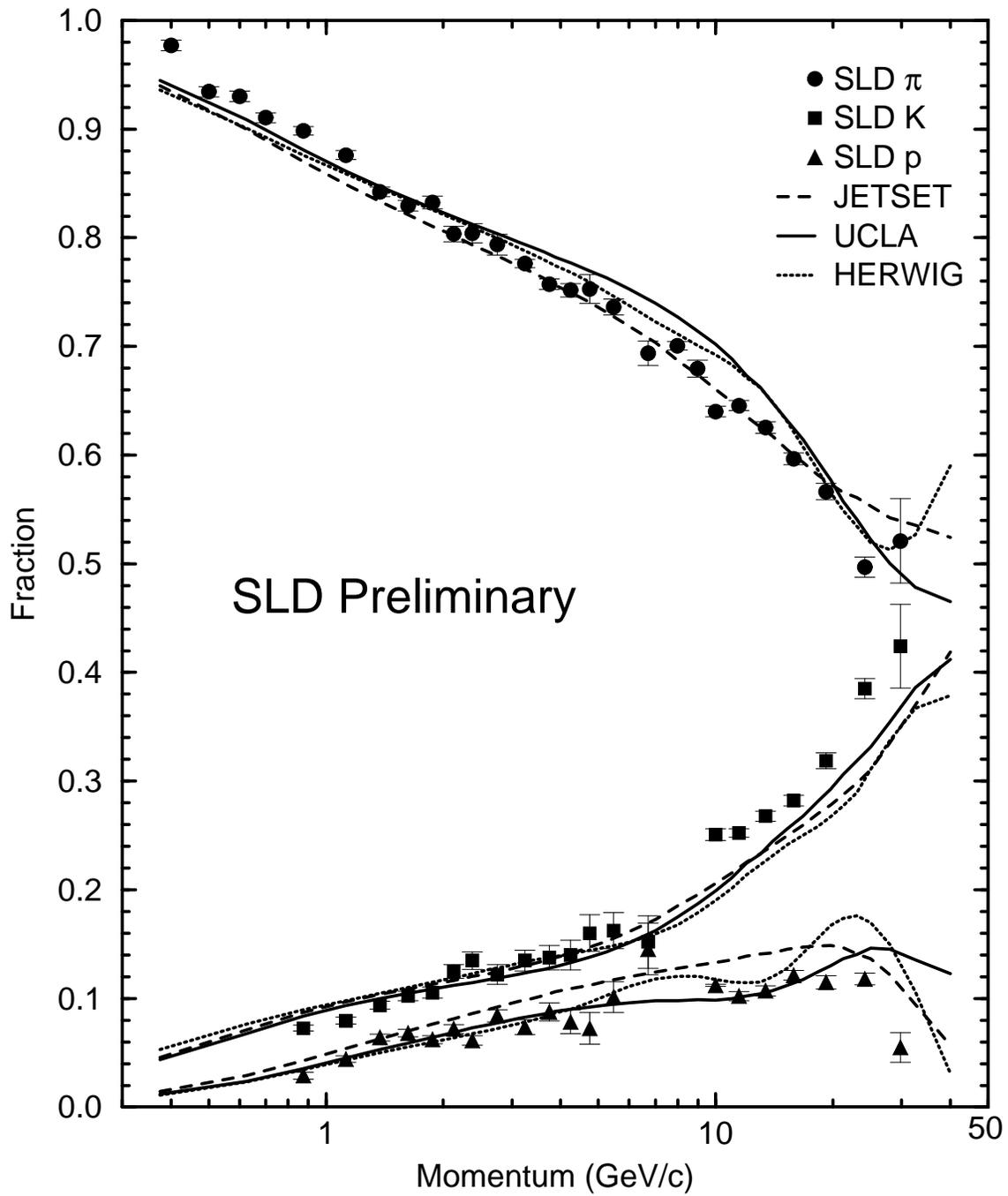}}\end{center}
  \caption{ 
 \label{ffruds}
Comparison of our charged hadron fractions in light-flavor events
with the predictions of three fragmentation models.
}
\end{figure}

In fig.~\ref{ffruds} we compare our measured charged hadron fractions in
light-flavor flavor events with the predictions of the three fragmentation
models. 
Qualitatively there is little difference between these data and those for the
inclusive sample (fig. \ref{fraxg}),
however these are more relevant for comparison with QCD predictions
based on the assumption of massless primary quark production,
as well as for determining parameters in fragmentation models.
We observe the same general differences between the predictions of the three
fragmentation models and the data as were seen above in the flavor-inclusive
sample.
This indicates that these deficiencies are in the fragmentation simulation
and not simply in the modelling of heavy hadron production and decay.

\begin{figure}
\vspace{-1.cm}
 \hspace*{0.5cm}   
   \epsfxsize=6.0in
   \begin{center}\mbox{\epsffile{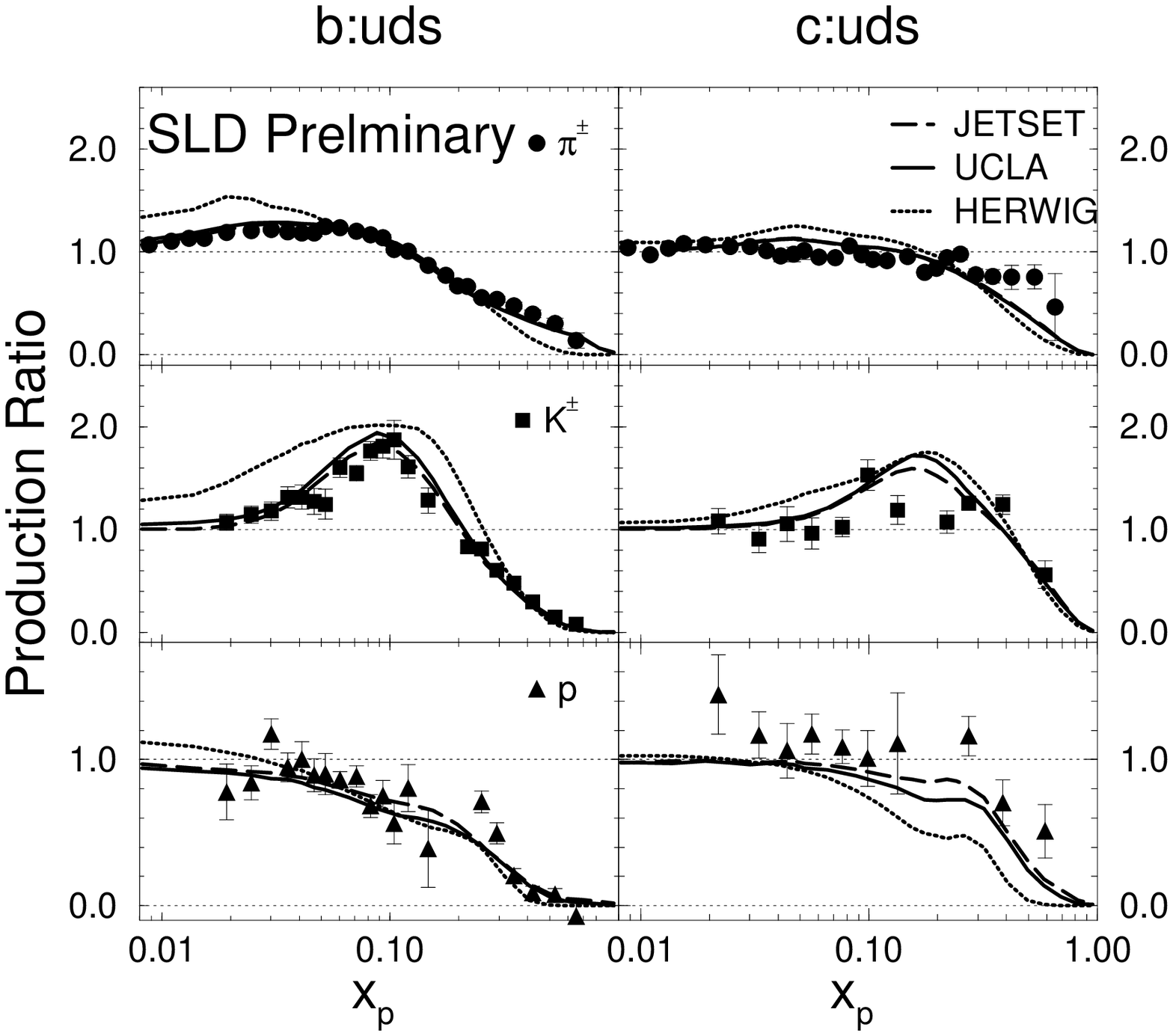}}\end{center}
  \caption{ 
 \label{xsrbumc}
Ratios of production rates in $b$-flavor events to those in light-flavor
events, along with the predictions of three fragmentation models.
   }
\end{figure}

In fig.~\ref{xsrbumc} we show the ratios of production in
$b$- to light-flavor and $c$- to light-flavor events for the three species.
The systematic errors on the particle identification largely cancel in
these ratios, and the resulting errors are predominantly statistical.
There is greater production of charged pions 
in $b$-flavor events at low momentum, with an approximately constant ratio for
$0.02<x_p<0.07$.
The production charged kaons is approximately equal
in the two samples at $x_p=0.02$, but the relative production in $b$-flavor
events then increases with $x_p$, peaking at $x_p \approx 0.07$.
There is approximately equal production of protons in $b$-flavor and
light-flavor events below $x_p=0.15$.
For $x_p>0.1$, production of all these particle species
falls faster with increasing momentum in $b$-flavor events.
These features are consistent with expectations based on the known properties
of \zbb\ events, namely that a large fraction of the event energy is carried by
the leading $B$- and $\bar{B}$-hadrons, which decay into a large number of
lighter particles.
Also shown in fig. \ref{xsrbumc} are the
predictions of the three fragmentation models, which reproduce these
features qualitatively, although 
HERWIG overestimates the pion and kaon ratios by a large factor at low $x_p$.

There is higher kaon production in $c$-flavor events than in light-flavor events
at $x_p \sim 0.1$, reflecting the tendency of $c$-jets to produce a fairly hard
charmed hadron whose decay products include a kaon carrying a large fraction of
its momentum.
There are fewer additional charged pions produced in $D$ decays than in $B$
decays, so that pion production is only slightly higher
in $c$-flavor events at very small $x_p$.
The pion $c$:light ratio starts to cut off at a larger value, $x_p \approx 0.3$,
than the corresponding $b$:light ratio, attributable to the lower average decay
multiplicity and softer fragmentation function of $D$ hadrons,
and the kaon and proton ratios are consistent with this cutoff point.
Again, all three fragmentation models reproduce the data qualitatively,
although HERWIG overestimates the pion ratio at small $x_p$, as it did in the
$b$:light case, and underestimates the proton ratio is large $x_p$.

\section{Leading Particle Effects}

We extended \cite{lpprl} these studies to look for differences between
particle and antiparticle production in quark (rather than antiquark) jets,
in order to address the question of whether e.g. 
a primary $u$-initiated jet contains more particles that contain a valence
$u$-quark
(e.g. $\pi^+$, $K^+$, p) than particles that do not
(e.g. $\pi^-$, $K^-$, $\bar{\rm p}$).
To this end we used the light
quark- and antiquark-tagged hemispheres described in section 2.

We measured the production rates per light quark jet
\begin{eqnarray}
R^{q}_{h} &=& {1\over{2N_{evts}}}{d\over{dx_{p}}}\left[ N(q\rightarrow
h)+N(\bar{q}\rightarrow\bar{h})\right],\\
R^{q}_{\bar{h}} &=& {1\over{2N_{evts}}}{d\over{dx_{p}}}\left[
N(q\rightarrow\bar{h})+N(\bar{q}\rightarrow h)\right],
\end{eqnarray}
where: $q$ and $\bar{q}$ represent light-flavor quark and antiquark jets
respectively; $N_{evts}$ is the total number of events in the sample; $h$
represents any of the identified hadrons $\pi^{-}$, $K^{-}$,
and p, and $\bar{h}$ indicates the
corresponding antiparticle.
Then, for example, $N(q\rightarrow h)$ is
the number of hadrons of type $h$ in light quark jets.

The charged hadron fractions analysis was repeated separately on the positively
and negatively charged tracks in each of the quark- and antiquark-tagged
samples.  Results for the positively charged tracks in the quark-tagged sample
and the negatively-charged tracks in the antiquark-tagged sample were
consistent, so these two samples were combined and labelled as positively
charged hadrons from light quark jets, yielding measured values of
$R^{q}_{\pi^{+}}$, $R^{q}_{K^{+}}$, and $R^{q}_{\rm p}$ in the tagged samples.
The same procedure applied to the remaining tracks yielded $R^{q}_{\pi^{-}}$,
$R^{q}_{K^{-}}$, and $R^{q}_{\bar{\rm p}}$.

It is essential to understand the contributions to these rates from
heavy-flavor events, which are typically large in the momentum range we cover
and show substantial differences between hadron and
antihadron due to decay products of the heavy hadrons.
This motivated our use of light-tagged events, and the residual heavy flavor
contributions were estimated from the simulation to be typically 15\% of the
observed hadrons.
This estimate was applied as a correction, yielding \pnb s
per light-quark-tagged jet.
The effect of this correction on the results was
negligible compared with the statistical errors.

For each hadron type, \pnb s in light quark jets were then extracted by
correcting for the light-tag bias and unfolding for the effective quark (vs.
antiquark) purity.
The purity was estimated from the simulation to be 0.72, which is slightly
lower than the value of 0.73 noted in section 2, reflecting the 
cutoff in acceptance of the barrel CRID at $|\cos\theta|=0.68$.

\begin{figure}
\vspace{-1.cm}
 \hspace*{0.5cm}   
   \epsfxsize=6.7in
   \begin{center}\mbox{\epsffile{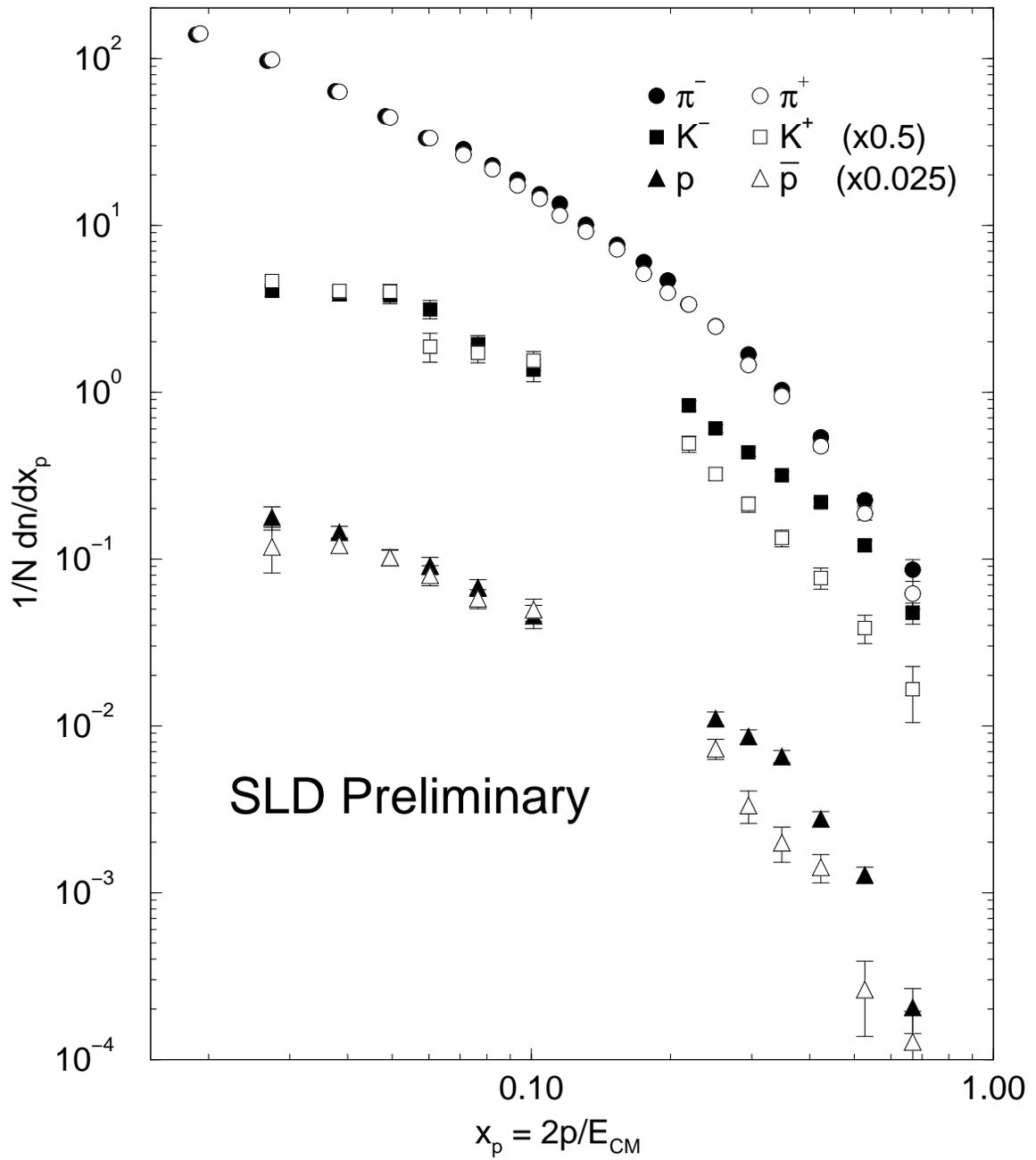}}\end{center}
  \caption{ 
 \label{xsqq}
Scaled momentum distributions of identified particles and their antiparticles
per light quark ($u$, $d$, $s$) jet.
    }
\end{figure}

The measured \pnb s per light quark jet are shown in fig \ref{xsqq}.
The errors shown are are the sum in quadrature of statistical errors and
those systematic errors arising from uncertainties in the heavy-flavor
background correction and the effective quark purity; 
the statistical errors dominate this total.
Systematic errors common to hadron and antihadron, such as those due to their
identification efficiencies, are not included,

\begin{figure}
\vspace{-1.cm}
 \hspace*{0.5cm}   
   \epsfxsize=5.9in
   \begin{center}\mbox{\epsffile{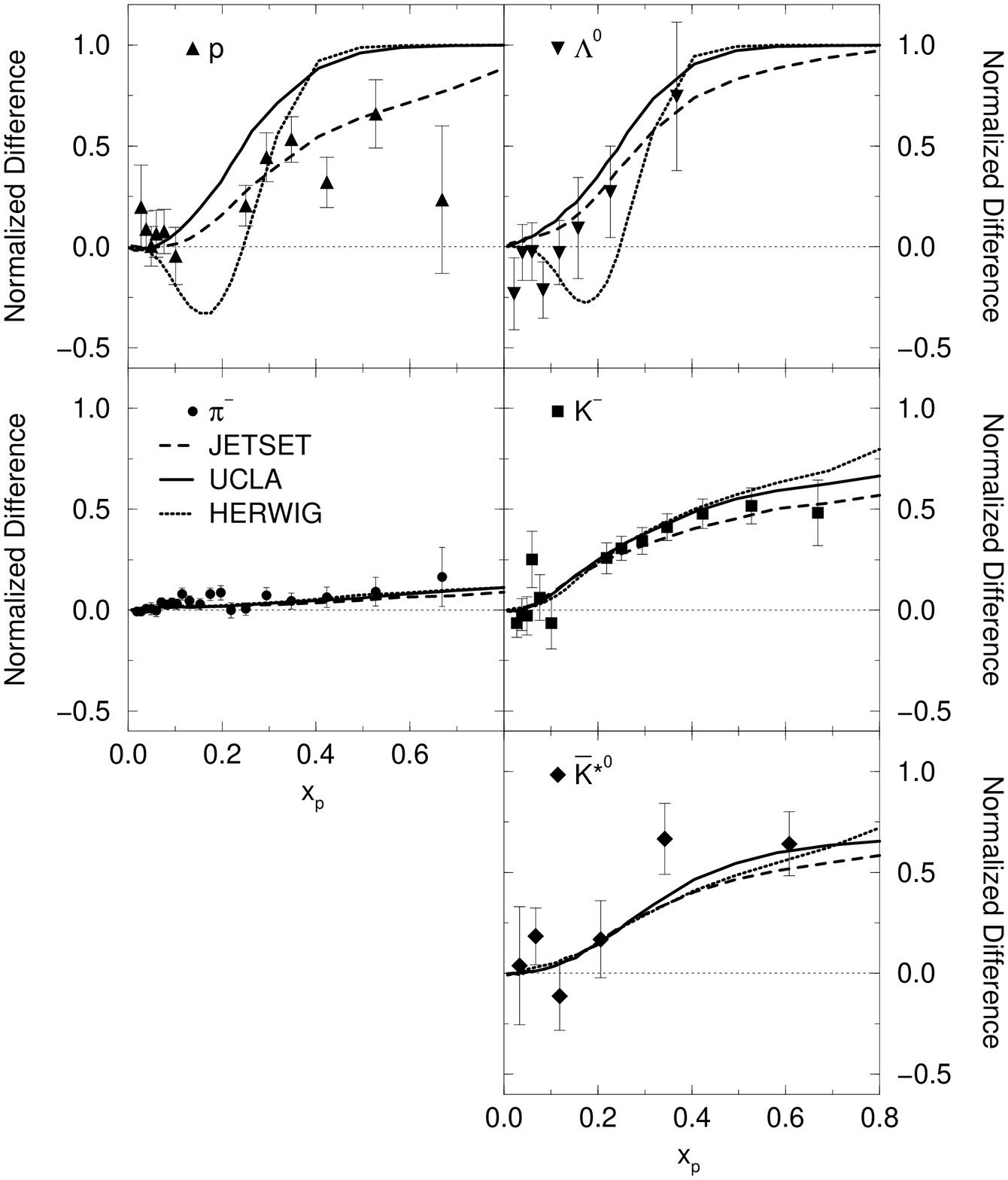}}\end{center}
  \caption{ 
 \label{ndqq}
Normalized production differences between hadrons and their
respective antihadrons in light quark jets.
Also shown are the predictions of three fragmentation models.
  }
\end{figure}

In all cases the hadron and antihadron \pnb s are consistent at low $x_p$.
For charged pions there are small differences at high $x_p$, and for the
other particles there are substantial differences, all of which appear to
increase with increasing $x_p$.
It is convenient to show these data in the form of the difference between hadron
and antihadron \pnb s normalized by the sum:
\begin{equation}
D_{h} =  {R^{q}_{h} - R^{q}_{\overline{h}}\over
          R^{q}_{h} + R^{q}_{\overline{h}}},
\end{equation}
The common systematic errors cancel explicitly in this variable.
Results are shown in fig \ref{ndqq}, along with our previous \cite{bfp} similar
results for the strange vector meson $K^{*0}$ and the $\Lambda^0$ hyperon.
A value of zero corresponds to equal production of hadron and antihadron,
and the data are consistent with zero at low $x_p$.
A value of $+1$ (--1) corresponds to complete dominance of (anti)hadrons $h$.

The baryon results are most straightforward to interpret.
Since baryons contain valence quarks and not antiquarks,
the excess of baryons over antibaryons in light quarks jets provides clear
evidence for the production of leading baryons at high scaled momentum.
The data suggest that the effect increases with $x_p$.

The interpretation for the mesons is more complicated,
since they contain one valence quark along with one antiquark.
All down-type quarks are produced equally and with the same SM 
forward-backward asymmetry in $Z^0$ decays, so that if a leading neutral
particle such as $K^{*0}$ ($d\bar{s}$) were produced equally in $d$ and
$\bar{s}$ jets
then one would observe $D_{\overline{K}^{*0}}=0$.
In the case of charged mesons such as $\pi^-$ ($d\bar{u}$),
the different production rates and forward-backward asymmetries of up- and
down-type quarks cause a nonzero dilution of leading particle effects.
At the $Z^0$, equal leading pion production in $u$- and $d$-jets
would lead to a dilution factor of 0.27.

Our measured $D_{\pi^-}$ are consistently above zero at high $x_p$, and 
consistently below 0.27$D_{\rm p}$, although statistically consistent at each
point with both.
This suggests that leading primary pions are produced,
but indicates that nonleading production of pions must be relatively large.
This could be due to a very soft leading pion momentum
distribution and/or a large ``background" contribution from decays of $\rho^0$,
$K^*$, etc.
Our measured $D_{K^-}$ are well above both zero and 0.27$D_{\rm p}$ for
$x_p > 0.2$.
This indicates
both substantial production of leading $K^{\pm}$ mesons at
high momentum, and a depletion of leading kaon
production in $u\bar{u}$ and $d\bar{d}$ events relative to $s\bar{s}$ events.

Assuming these high-momentum kaons to be directly produced in the fragmentation
process, this amounts to a direct observation of a suppression of $s\bar{s}$
production from the vacuum with respect to $u\bar{u}$ or $d\bar{d}$ production.
Assuming {\it all} $K^{\pm}$ in the
range $x_{p}>0.5$ to be leading, we calculate
$\gamma_s = 0.26\pm 0.06$,
consistent with values~\cite{saxon} derived from inclusive measurements of the
relative production rates of strange and non-strange, pseudoscalar and vector
mesons.

Also shown in fig. \ref{ndqq} are the predictions of the three Fragmentation
models.  All three are consistent with the meson data and with the $\Lambda^0$
data.  The JETSET model is also consistent with the proton data, however the
other two models predict a saturated value of $D_{\rm p}$ for $x_p>0.4$ that is
inconsistent with the data.

\section{Summary and Conclusions}

   Using the SLD Cherenkov Ring Imaging Detector we have made preliminary
measurements of charged pion, kaon and proton
production over most of the momentum range in hadronic $Z^0$ decays.
We find the predictions of the JETSET, UCLA and HERWIG fragmentation models to
be in qualitative agreement with our data.
These results are in agreement with those from
previous experiments.

By isolating high-purity light- and $b$-flavor samples, we have
measured hadron production in light-flavor events, as well
as in $c$- and $b$-flavor events.
We find substantial differences in particle production between light- and
heavy-flavor events, with the latter producing more mesons overall, but far
fewer at high momentum.
These qualitative features are expected given the hard fragmentation and high
average decay multiplicity of heavy hadrons.
The light-flavor sample is more suitable for testing predictions of QCD that
assume massless quarks, as well as for testing fragmentation models.
We find differences between fragmentation model predictions and our data similar
to those found in the inclusive sample, indicating that the deficiencies
lie in the
simulation of fragmentation rather than in that of heavy hadron production and
decay.


By isolating high-purity light-quark and light-antiquark samples, we have
made the first comparison of hadron and antihadron production in light-quark
jets in $e^+e^-$ annihilation.
We observed an excess of p over $\bar{\rm p}$, which appears to
increase with momentum, and provides direct evidence for the ``leading
particle'' hypothesis that high momentum protons are more likely to contain the
primary quark.
We also observed a large excess of high momentum $K^-$
over $K^+$ indicating that a high momentum kaon
is likely to contain a primary quark or antiquark from the $Z^0$ decay, and
that leading kaons are produced predominantly in $s\bar{s}$ events rather than
$d\bar{d}$ or $u\bar{u}$ events.
We observe only a small excess of $\pi^-$ over $\pi^+$ at high momentum, due in
part to the cancellation of the signal from $u\bar{u}$ and $d\bar{d}$ events,
but also suggesting a large nonleading pion fraction even in this momentum
region.


\section*{Acknowledgements}
We thank the personnel of the SLAC accelerator department and the
technical
staffs of our collaborating institutions for their outstanding efforts
on our behalf.

\vskip .5truecm

\vbox{
\noindent
 $^*$This work was supported by Department of Energy
  contracts:
  DE-FG02-91ER40676 (BU),
  DE-FG03-91ER40618 (UCSB),
  DE-FG03-92ER40689 (UCSC),
  DE-FG03-93ER40788 (CSU),
  DE-FG02-91ER40672 (Colorado),
  DE-FG02-91ER40677 (Illinois),
  DE-AC03-76SF00098 (LBL),
  DE-FG02-92ER40715 (Massachusetts),
  DE-FC02-94ER40818 (MIT),
  DE-FG03-96ER40969 (Oregon),
  DE-AC03-76SF00515 (SLAC),
  DE-FG05-91ER40627 (Tennessee),
  DE-FG02-95ER40896 (Wisconsin),
  DE-FG02-92ER40704 (Yale);
  National Science Foundation grants:
  PHY-91-13428 (UCSC),
  PHY-89-21320 (Columbia),
  PHY-92-04239 (Cincinnati),
  PHY-95-10439 (Rutgers),
  PHY-88-19316 (Vanderbilt),
  PHY-92-03212 (Washington);
  The UK Particle Physics and Astronomy Research Council
  (Brunel, Oxford and RAL);
  The Istituto Nazionale di Fisica Nucleare of Italy
  (Bologna, Ferrara, Frascati, Pisa, Padova, Perugia);
  The Japan-US Cooperative Research Project on High Energy Physics
  (Nagoya, Tohoku);
  The Korea Science and Engineering Foundation (Soongsil).}

\section*{$^{**}$List of Authors}

%
%
%
\begin{center}
\def\iADEL{$^{(1)}$}
\def\iAOMORI{$^{(2)}$}
\def\iBOLO{$^{(3)}$}
\def\iBRI{$^{(4)}$}
\def\iBRUN{$^{(5)}$}
\def\iBU{$^{(6)}$}
\def\iCINC{$^{(7)}$}
\def\iCOLO{$^{(8)}$}
\def\iCOLU{$^{(9)}$}
\def\iCSU{$^{(10)}$}
\def\iFERR{$^{(11)}$}
\def\iFRAS{$^{(12)}$}
\def\iILLI{$^{(13)}$}
\def\iJHU{$^{(14)}$}
\def\iLBL{$^{(15)}$}
\def\iLTU{$^{(16)}$}
\def\iMASS{$^{(17)}$}
\def\iMISSI{$^{(18)}$}
\def\iMIT{$^{(19)}$}
\def\iMOSCOW{$^{(20)}$}
\def\iNAGO{$^{(21)}$}
\def\iOREG{$^{(22)}$}
\def\iOXF{$^{(23)}$}
\def\iPADO{$^{(24)}$}
\def\iPERU{$^{(25)}$}
\def\iPISA{$^{(26)}$}
\def\iRAL{$^{(27)}$}
\def\iRUTG{$^{(28)}$}
\def\iSLAC{$^{(29)}$}
\def\iSOGA{$^{(30)}$}
\def\iSOONG{$^{(31)}$}
\def\iTENN{$^{(32)}$}
\def\iTOHO{$^{(33)}$}
\def\iUCSB{$^{(34)}$}
\def\iUCSC{$^{(35)}$}
\def\iUVIC{$^{(36)}$}
\def\iVAND{$^{(37)}$}
\def\iWASH{$^{(38)}$}
\def\iWISC{$^{(39)}$}
\def\iYALE{$^{(40)}$}

  \baselineskip=.75\baselineskip  
\mbox{Kenji  Abe\unskip,\iNAGO}
\mbox{Koya Abe\unskip,\iTOHO}
\mbox{T. Abe\unskip,\iSLAC}
\mbox{I.Adam\unskip,\iSLAC}
\mbox{T.  Akagi\unskip,\iSLAC}
\mbox{N. J. Allen\unskip,\iBRUN}
\mbox{W.W. Ash\unskip,\iSLAC}
\mbox{D. Aston\unskip,\iSLAC}
\mbox{K.G. Baird\unskip,\iMASS}
\mbox{C. Baltay\unskip,\iYALE}
\mbox{H.R. Band\unskip,\iWISC}
\mbox{M.B. Barakat\unskip,\iLTU}
\mbox{O. Bardon\unskip,\iMIT}
\mbox{T.L. Barklow\unskip,\iSLAC}
\mbox{G. L. Bashindzhagyan\unskip,\iMOSCOW}
\mbox{J.M. Bauer\unskip,\iMISSI}
\mbox{G. Bellodi\unskip,\iOXF}
\mbox{R. Ben-David\unskip,\iYALE}
\mbox{A.C. Benvenuti\unskip,\iBOLO}
\mbox{G.M. Bilei\unskip,\iPERU}
\mbox{D. Bisello\unskip,\iPADO}
\mbox{G. Blaylock\unskip,\iMASS}
\mbox{J.R. Bogart\unskip,\iSLAC}
\mbox{G.R. Bower\unskip,\iSLAC}
\mbox{J. E. Brau\unskip,\iOREG}
\mbox{M. Breidenbach\unskip,\iSLAC}
\mbox{W.M. Bugg\unskip,\iTENN}
\mbox{D. Burke\unskip,\iSLAC}
\mbox{T.H. Burnett\unskip,\iWASH}
\mbox{P.N. Burrows\unskip,\iOXF}
\mbox{A. Calcaterra\unskip,\iFRAS}
\mbox{D. Calloway\unskip,\iSLAC}
\mbox{B. Camanzi\unskip,\iFERR}
\mbox{M. Carpinelli\unskip,\iPISA}
\mbox{R. Cassell\unskip,\iSLAC}
\mbox{R. Castaldi\unskip,\iPISA}
\mbox{A. Castro\unskip,\iPADO}
\mbox{M. Cavalli-Sforza\unskip,\iUCSC}
\mbox{A. Chou\unskip,\iSLAC}
\mbox{E. Church\unskip,\iWASH}
\mbox{H.O. Cohn\unskip,\iTENN}
\mbox{J.A. Coller\unskip,\iBU}
\mbox{M.R. Convery\unskip,\iSLAC}
\mbox{V. Cook\unskip,\iWASH}
\mbox{R. Cotton\unskip,\iBRUN}
\mbox{R.F. Cowan\unskip,\iMIT}
\mbox{D.G. Coyne\unskip,\iUCSC}
\mbox{G. Crawford\unskip,\iSLAC}
\mbox{C.J.S. Damerell\unskip,\iRAL}
\mbox{M. N. Danielson\unskip,\iCOLO}
\mbox{M. Daoudi\unskip,\iSLAC}
\mbox{N. de Groot\unskip,\iBRI}
\mbox{R. Dell'Orso\unskip,\iPERU}
\mbox{P.J. Dervan\unskip,\iBRUN}
\mbox{R. de Sangro\unskip,\iFRAS}
\mbox{M. Dima\unskip,\iCSU}
\mbox{A. D'Oliveira\unskip,\iCINC}
\mbox{D.N. Dong\unskip,\iMIT}
\mbox{M. Doser\unskip,\iSLAC}
\mbox{R. Dubois\unskip,\iSLAC}
\mbox{B.I. Eisenstein\unskip,\iILLI}
\mbox{V. Eschenburg\unskip,\iMISSI}
\mbox{E. Etzion\unskip,\iWISC}
\mbox{S. Fahey\unskip,\iCOLO}
\mbox{D. Falciai\unskip,\iFRAS}
\mbox{C. Fan\unskip,\iCOLO}
\mbox{J.P. Fernandez\unskip,\iUCSC}
\mbox{M.J. Fero\unskip,\iMIT}
\mbox{K.Flood\unskip,\iMASS}
\mbox{R. Frey\unskip,\iOREG}
\mbox{J. Gifford\unskip,\iUVIC}
\mbox{T. Gillman\unskip,\iRAL}
\mbox{G. Gladding\unskip,\iILLI}
\mbox{S. Gonzalez\unskip,\iMIT}
\mbox{E. R. Goodman\unskip,\iCOLO}
\mbox{E.L. Hart\unskip,\iTENN}
\mbox{J.L. Harton\unskip,\iCSU}
\mbox{A. Hasan\unskip,\iBRUN}
\mbox{K. Hasuko\unskip,\iTOHO}
\mbox{S. J. Hedges\unskip,\iBU}
\mbox{S.S. Hertzbach\unskip,\iMASS}
\mbox{M.D. Hildreth\unskip,\iSLAC}
\mbox{J. Huber\unskip,\iOREG}
\mbox{M.E. Huffer\unskip,\iSLAC}
\mbox{E.W. Hughes\unskip,\iSLAC}
\mbox{X.Huynh\unskip,\iSLAC}
\mbox{H. Hwang\unskip,\iOREG}
\mbox{M. Iwasaki\unskip,\iOREG}
\mbox{D. J. Jackson\unskip,\iRAL}
\mbox{P. Jacques\unskip,\iRUTG}
\mbox{J.A. Jaros\unskip,\iSLAC}
\mbox{Z.Y. Jiang\unskip,\iSLAC}
\mbox{A.S. Johnson\unskip,\iSLAC}
\mbox{J.R. Johnson\unskip,\iWISC}
\mbox{R.A. Johnson\unskip,\iCINC}
\mbox{T. Junk\unskip,\iSLAC}
\mbox{R. Kajikawa\unskip,\iNAGO}
\mbox{M. Kalelkar\unskip,\iRUTG}
\mbox{Y. Kamyshkov\unskip,\iTENN}
\mbox{H.J. Kang\unskip,\iRUTG}
\mbox{I. Karliner\unskip,\iILLI}
\mbox{H. Kawahara\unskip,\iSLAC}
\mbox{Y. D. Kim\unskip,\iSOGA}
\mbox{M.E. King\unskip,\iSLAC}
\mbox{R. King\unskip,\iSLAC}
\mbox{R.R. Kofler\unskip,\iMASS}
\mbox{N.M. Krishna\unskip,\iCOLO}
\mbox{R.S. Kroeger\unskip,\iMISSI}
\mbox{M. Langston\unskip,\iOREG}
\mbox{A. Lath\unskip,\iMIT}
\mbox{D.W.G. Leith\unskip,\iSLAC}
\mbox{V. Lia\unskip,\iMIT}
\mbox{C.Lin\unskip,\iMASS}
\mbox{M.X. Liu\unskip,\iYALE}
\mbox{X. Liu\unskip,\iUCSC}
\mbox{M. Loreti\unskip,\iPADO}
\mbox{A. Lu\unskip,\iUCSB}
\mbox{H.L. Lynch\unskip,\iSLAC}
\mbox{J. Ma\unskip,\iWASH}
\mbox{G. Mancinelli\unskip,\iRUTG}
\mbox{S. Manly\unskip,\iYALE}
\mbox{G. Mantovani\unskip,\iPERU}
\mbox{T.W. Markiewicz\unskip,\iSLAC}
\mbox{T. Maruyama\unskip,\iSLAC}
\mbox{H. Masuda\unskip,\iSLAC}
\mbox{E. Mazzucato\unskip,\iFERR}
\mbox{A.K. McKemey\unskip,\iBRUN}
\mbox{B.T. Meadows\unskip,\iCINC}
\mbox{G. Menegatti\unskip,\iFERR}
\mbox{R. Messner\unskip,\iSLAC}
\mbox{P.M. Mockett\unskip,\iWASH}
\mbox{K.C. Moffeit\unskip,\iSLAC}
\mbox{T.B. Moore\unskip,\iYALE}
\mbox{M.Morii\unskip,\iSLAC}
\mbox{D. Muller\unskip,\iSLAC}
\mbox{V.Murzin\unskip,\iMOSCOW}
\mbox{T. Nagamine\unskip,\iTOHO}
\mbox{S. Narita\unskip,\iTOHO}
\mbox{U. Nauenberg\unskip,\iCOLO}
\mbox{H. Neal\unskip,\iSLAC}
\mbox{M. Nussbaum\unskip,\iCINC}
\mbox{N.Oishi\unskip,\iNAGO}
\mbox{D. Onoprienko\unskip,\iTENN}
\mbox{L.S. Osborne\unskip,\iMIT}
\mbox{R.S. Panvini\unskip,\iVAND}
\mbox{C. H. Park\unskip,\iSOONG}
\mbox{T.J. Pavel\unskip,\iSLAC}
\mbox{I. Peruzzi\unskip,\iFRAS}
\mbox{M. Piccolo\unskip,\iFRAS}
\mbox{L. Piemontese\unskip,\iFERR}
\mbox{K.T. Pitts\unskip,\iOREG}
\mbox{R.J. Plano\unskip,\iRUTG}
\mbox{R. Prepost\unskip,\iWISC}
\mbox{C.Y. Prescott\unskip,\iSLAC}
\mbox{G.D. Punkar\unskip,\iSLAC}
\mbox{J. Quigley\unskip,\iMIT}
\mbox{B.N. Ratcliff\unskip,\iSLAC}
\mbox{T.W. Reeves\unskip,\iVAND}
\mbox{J. Reidy\unskip,\iMISSI}
\mbox{P.L. Reinertsen\unskip,\iUCSC}
\mbox{P.E. Rensing\unskip,\iSLAC}
\mbox{L.S. Rochester\unskip,\iSLAC}
\mbox{P.C. Rowson\unskip,\iCOLU}
\mbox{J.J. Russell\unskip,\iSLAC}
\mbox{O.H. Saxton\unskip,\iSLAC}
\mbox{T. Schalk\unskip,\iUCSC}
\mbox{R.H. Schindler\unskip,\iSLAC}
\mbox{B.A. Schumm\unskip,\iUCSC}
\mbox{J. Schwiening\unskip,\iSLAC}
\mbox{S. Sen\unskip,\iYALE}
\mbox{V.V. Serbo\unskip,\iSLAC}
\mbox{M.H. Shaevitz\unskip,\iCOLU}
\mbox{J.T. Shank\unskip,\iBU}
\mbox{G. Shapiro\unskip,\iLBL}
\mbox{D.J. Sherden\unskip,\iSLAC}
\mbox{K. D. Shmakov\unskip,\iTENN}
\mbox{C. Simopoulos\unskip,\iSLAC}
\mbox{N.B. Sinev\unskip,\iOREG}
\mbox{S.R. Smith\unskip,\iSLAC}
\mbox{M. B. Smy\unskip,\iCSU}
\mbox{J.A. Snyder\unskip,\iYALE}
\mbox{H. Staengle\unskip,\iCSU}
\mbox{A. Stahl\unskip,\iSLAC}
\mbox{P. Stamer\unskip,\iRUTG}
\mbox{H. Steiner\unskip,\iLBL}
\mbox{R. Steiner\unskip,\iADEL}
\mbox{M.G. Strauss\unskip,\iMASS}
\mbox{D. Su\unskip,\iSLAC}
\mbox{F. Suekane\unskip,\iTOHO}
\mbox{A. Sugiyama\unskip,\iNAGO}
\mbox{S. Suzuki\unskip,\iNAGO}
\mbox{M. Swartz\unskip,\iJHU}
\mbox{A. Szumilo\unskip,\iWASH}
\mbox{T. Takahashi\unskip,\iSLAC}
\mbox{F.E. Taylor\unskip,\iMIT}
\mbox{J. Thom\unskip,\iSLAC}
\mbox{E. Torrence\unskip,\iMIT}
\mbox{N. K. Toumbas\unskip,\iSLAC}
\mbox{T. Usher\unskip,\iSLAC}
\mbox{C. Vannini\unskip,\iPISA}
\mbox{J. Va'vra\unskip,\iSLAC}
\mbox{E. Vella\unskip,\iSLAC}
\mbox{J.P. Venuti\unskip,\iVAND}
\mbox{R. Verdier\unskip,\iMIT}
\mbox{P.G. Verdini\unskip,\iPISA}
\mbox{D. L. Wagner\unskip,\iCOLO}
\mbox{S.R. Wagner\unskip,\iSLAC}
\mbox{A.P. Waite\unskip,\iSLAC}
\mbox{S. Walston\unskip,\iOREG}
\mbox{J.Wang\unskip,\iSLAC}
\mbox{S.J. Watts\unskip,\iBRUN}
\mbox{A.W. Weidemann\unskip,\iTENN}
\mbox{E. R. Weiss\unskip,\iWASH}
\mbox{J.S. Whitaker\unskip,\iBU}
\mbox{S.L. White\unskip,\iTENN}
\mbox{F.J. Wickens\unskip,\iRAL}
\mbox{B. Williams\unskip,\iCOLO}
\mbox{D.C. Williams\unskip,\iMIT}
\mbox{S.H. Williams\unskip,\iSLAC}
\mbox{S. Willocq\unskip,\iMASS}
\mbox{R.J. Wilson\unskip,\iCSU}
\mbox{W.J. Wisniewski\unskip,\iSLAC}
\mbox{J. L. Wittlin\unskip,\iMASS}
\mbox{M. Woods\unskip,\iSLAC}
\mbox{G.B. Word\unskip,\iVAND}
\mbox{T.R. Wright\unskip,\iWISC}
\mbox{J. Wyss\unskip,\iPADO}
\mbox{R.K. Yamamoto\unskip,\iMIT}
\mbox{J.M. Yamartino\unskip,\iMIT}
\mbox{X. Yang\unskip,\iOREG}
\mbox{J. Yashima\unskip,\iTOHO}
\mbox{S.J. Yellin\unskip,\iUCSB}
\mbox{C.C. Young\unskip,\iSLAC}
\mbox{H. Yuta\unskip,\iAOMORI}
\mbox{G. Zapalac\unskip,\iWISC}
\mbox{R.W. Zdarko\unskip,\iSLAC}
\mbox{J. Zhou\unskip.\iOREG}

\it
  \vskip \baselineskip                   
  \vskip \baselineskip        
  \baselineskip=.75\baselineskip   
\iADEL
  Adelphi University, Garden City, New York 11530, \break
\iAOMORI
  Aomori University, Aomori , 030 Japan, \break
\iBOLO
  INFN Sezione di Bologna, I-40126, Bologna Italy, \break
\iBRI
  University of Bristol, Bristol, U.K., \break
\iBRUN
  Brunel University, Uxbridge, Middlesex, UB8 3PH United Kingdom, \break
\iBU
  Boston University, Boston, Massachusetts 02215, \break
\iCINC
  University of Cincinnati, Cincinnati, Ohio 45221, \break
\iCOLO
  University of Colorado, Boulder, Colorado 80309, \break
\iCOLU
  Columbia University, New York, New York 10533, \break
\iCSU
  Colorado State University, Ft. Collins, Colorado 80523, \break
\iFERR
  INFN Sezione di Ferrara and Universita di Ferrara, I-44100 Ferrara, Italy, \break
\iFRAS
  INFN Lab. Nazionali di Frascati, I-00044 Frascati, Italy, \break
\iILLI
  University of Illinois, Urbana, Illinois 61801, \break
\iJHU
  Johns Hopkins University, Baltimore, MD 21218-2686, \break
\iLBL
  Lawrence Berkeley Laboratory, University of California, Berkeley, California 94720, \break
\iLTU
  Louisiana Technical University - Ruston,LA 71272, \break
\iMASS
  University of Massachusetts, Amherst, Massachusetts 01003, \break
\iMISSI
  University of Mississippi, University, Mississippi 38677, \break
\iMIT
  Massachusetts Institute of Technology, Cambridge, Massachusetts 02139, \break
\iMOSCOW
  Institute of Nuclear Physics, Moscow State University, 119899, Moscow Russia, \break
\iNAGO
  Nagoya University, Chikusa-ku, Nagoya 464 Japan, \break
\iOREG
  University of Oregon, Eugene, Oregon 97403, \break
\iOXF
  Oxford University, Oxford, OX1 3RH, United Kingdom, \break
\iPADO
  INFN Sezione di Padova and Universita di Padova I-35100, Padova, Italy, \break
\iPERU
  INFN Sezione di Perugia and Universita di Perugia, I-06100 Perugia, Italy, \break
\iPISA
  INFN Sezione di Pisa and Universita di Pisa, I-56010 Pisa, Italy, \break
\iRAL
  Rutherford Appleton Laboratory, Chilton, Didcot, Oxon OX11 0QX United Kingdom, \break
\iRUTG
  Rutgers University, Piscataway, New Jersey 08855, \break
\iSLAC
  Stanford Linear Accelerator Center, Stanford University, Stanford, California 94309, \break
\iSOGA
  Sogang University, Seoul, Korea, \break
\iSOONG
  Soongsil University, Seoul, Korea 156-743, \break
\iTENN
  University of Tennessee, Knoxville, Tennessee 37996, \break
\iTOHO
  Tohoku University, Sendai 980, Japan, \break
\iUCSB
  University of California at Santa Barbara, Santa Barbara, California 93106, \break
\iUCSC
  University of California at Santa Cruz, Santa Cruz, California 95064, \break
\iUVIC
  University of Victoria, Victoria, B.C., Canada, V8W 3P6, \break
\iVAND
  Vanderbilt University, Nashville,Tennessee 37235, \break
\iWASH
  University of Washington, Seattle, Washington 98105, \break
\iWISC
  University of Wisconsin, Madison,Wisconsin 53706, \break
\iYALE
  Yale University, New Haven, Connecticut 06511. \break

\rm
%

\end{center}




\baselineskip=12pt 

\begin{thebibliography}{99}

\bibitem{mlla} T.I.~Azimov, Y.L.~Dokshitzer, V.A.~Khoze and S.I.~Troyan, Z.
Phys. {\bf C27} (1985) 65.

\bibitem{lphd} See e.g. I.G. Knowles and G.D. Lafferty, CERN-PPE/97-040.

\bibitem{sld} SLD Design Report, SLAC-Report 273, (1984).

\bibitem{sldalphas}
SLD Collaboration: K.~Abe et  al., Phys.\ Rev.\ {\bf D51} (1995) 962.

\bibitem{cdc}
M.D. Hildreth {\em et al.}, Nucl. Inst. Meth. {\bf A367} (1995) 111.

\bibitem{vxd}
C. J. S. Damerell {\em et al.}, Nucl. Inst. Meth. {\bf A288}~(1990)~236; \\
C. J. S. Damerell {\em et al.}, Nucl. Inst. Meth. {\bf A400}~(1997)~287.

\bibitem{crid}
K. Abe, {\em et al.}, Nucl. Inst. Meth. {\bf A343} (1994) 74.

\bibitem{bfp}
SLD Collab., K.~Abe {\em et al.}, Phys. Rev. {\bf D59} (1999) 52001.

\bibitem{thrust}
S. Brandt {\em et al.}, Phys. Lett. {\bf 12}~(1964)~57;\\
E. Farhi, Phys. Rev. Lett. {\bf 39}~(1977)~1587.

\bibitem {lac} 
D. Axen et al., Nucl. Inst. Meth. {\bf A328}~(1993)~472.

\bibitem{mikeh}
SLD Collab., K.~Abe {\em et al.}, Phys. Rev. {\bf D53} (1996) 2271.

\bibitem{jetset}
T. Sj\"ostrand, Comp. Phys. Comm. {\bf 82} (1994) 74.

\bibitem{tune}
P. N. Burrows, Z. Phys. {\bf C41} (1988) 375.\\
OPAL Collaboration, M.Z. Akrawy et al., Z. Phys. {\bf C47} (1990) 505.
 
\bibitem{sldsim}
SLD Collaboration, K. Abe et al., Phys. Rev. Lett. {\bf 79} (1997) 590.

\bibitem{geant}
R. Brun et al., Report No. CERN-DD/EE/84-1 (1989). 
 
\bibitem{davea}
K. Abe, {\it et al.}, Nucl. Inst. and Meth. {\bf A371} (1996) 195

\bibitem{delphi}
DELPHI Collab., P.~Abreu {\em et al.}, Nucl. Phys. {\bf B444} (1995) 3.

\bibitem{opal}
OPAL Collab., P.D.~Acton {\em et al.}, Z. Phys. {\bf C63}~(1994)~181.

\bibitem{aleph}
ALEPH Collab., D.~Buskulic {\em et al.}, Z. Phys. {\bf C66} (1995) 355.


\bibitem{ucla}
S. Chun and C. Buchanan, Phys. Rep. {\bf 292} (1998) 239.

\bibitem{herwig}
G. Marchesini {\em et al.}, Comp. Phys. Comm. {\bf 67}~(1992)~465.

\bibitem{tomp}
T.J. Pavel, Ph. D Thesis, Stanford University, January 1997;
SLAC-Report-495.



\bibitem{pdg}
Particle Data Group, Phys. Rev. {\bf D54} (1996) 1.

\bibitem{lpprl} SLD Collab., K. Abe {\em et al}.,
Phys. Rev. Lett. {\bf 78} (1997) 3442.

\bibitem{saxon}
D.H. Saxon, {\it High Energy Electron-Positron Physics}, Eds. A. Ali and P.
S\"oding, World Scientific (1988), p. 539.


\end{thebibliography}
\end{document}